\documentclass{article}

\usepackage[final]{neurips_2025}

\usepackage[utf8]{inputenc} 
\usepackage[T1]{fontenc}    
\usepackage{hyperref}       
\usepackage{url}            
\usepackage{booktabs}       
\usepackage{amsfonts}       
\usepackage{nicefrac}       
\usepackage{microtype}      
\usepackage{xcolor}         
\usepackage{bbm}
\usepackage{amsmath}
\usepackage{amssymb}
\usepackage{graphicx}
\usepackage{placeins}
\usepackage{subcaption}
\usepackage{enumitem}
\usepackage{algorithm}
\usepackage{algorithmic}

\usepackage{tikz}
\usetikzlibrary{
    shapes.geometric, 
    shapes.symbols,   
    arrows.meta,      
    positioning,      
    calc              
}
\usepackage{float}        
\usepackage{subcaption}   
\usepackage{caption}      

\usepackage{booktabs}
\usepackage{multirow}
\usepackage{xcolor}
\usepackage{colortbl}
\usepackage{graphicx}

\definecolor{best}{RGB}{0, 100, 0}      
\definecolor{good}{RGB}{34, 139, 34}    
\definecolor{medium}{RGB}{255, 215, 0}  
\definecolor{low}{RGB}{255, 140, 0}     

\usepackage{tikz}
\usetikzlibrary{shapes.geometric, arrows.meta, positioning, fit, calc, decorations.pathreplacing, calligraphy} 
\tikzstyle{block} = [rectangle, rounded corners, draw=black, fill=blue!20, text width=3cm, text centered, minimum height=1.5cm]
\tikzstyle{data} = [rectangle, draw=black, fill=green!20, text width=3cm, text centered, minimum height=1cm, font=\scriptsize]
\tikzstyle{io} = [trapezium, trapezium left angle=70, trapezium right angle=110, draw=black, fill=orange!30, text width=3cm, text centered, minimum height=1cm]
\tikzstyle{decision} = [diamond, draw=black, fill=red!20, text centered, minimum width=2.5cm, minimum height=1.5cm, aspect=1.5]
\tikzstyle{process} = [rectangle, draw=black, fill=yellow!30, text width=4.5cm, text centered, minimum height=1.5cm, align=center] 
\tikzstyle{llm} = [cylinder, shape border rotate=90, draw=black, fill=purple!20, text width=2.5cm, text centered, minimum height=1.5cm, aspect=0.5]
\tikzstyle{arrow} = [thick,->,>=Stealth]
\tikzstyle{comment} = [font=\tiny\itshape, text width=3cm, below left=0.1cm and 0.1cm of #1] 
\tikzstyle{brace} = [decorate, decoration={calligraphic brace, amplitude=4pt, mirror}, thick] 
\usepackage{tikz}
\usetikzlibrary{shapes,arrows,positioning}

\title{Benchmarking Large Language Models for Zero-shot and Few-shot Phishing URL Detection}

\author{%
  Najmul Hasan\\
  University of North Carolina at Pembroke\\
  \texttt{nh0033@bravemail.uncp.edu} \\
  \And
  Prashanth BusiReddyGari \\
  University of North Carolina at Pembroke\\
  \texttt{Prashanth.BusiReddyGari@uncp.edu} \\
}

\begin{document}

\maketitle

\begin{abstract}
\label{sec:abstract}

The Uniform Resource Locator (URL), introduced in a connectivity-first era to
define access and locate resources, remains historically limited, lacking future-
proof mechanisms for security, trust, or resilience against fraud and abuse, despite
the introduction of reactive protections like HTTPS during the cybersecurity era.
In the current AI-first threatscape, deceptive URLs have reached unprecedented
sophistication due to the widespread use of generative AI by cybercriminals and
the AI-vs-AI arms race to produce context-aware phishing websites and URLs
that are virtually indistinguishable to both users and traditional detection tools.
Although AI-generated phishing accounted for a small fraction of filter-bypassing
attacks in 2024, phishing volume has escalated over 4,000\% since 2022, with nearly
50\% more attacks evading detection. At the rate the threatscape is escalating, and
phishing tactics are emerging faster than labeled data can be produced, zero-shot
and few-shot learning with large language models (LLMs) offers a timely and
adaptable solution, enabling generalization with minimal supervision. Given the critical importance of phishing URL detection in large-scale cybersecurity defense systems, we present a comprehensive benchmark of LLMs under a unified zero-shot and few-shot prompting framework and reveal operational trade-offs. Our evaluation uses a balanced dataset with consistent prompts, offering detailed analysis of performance, generalization, and model efficacy, quantified by accuracy, precision, recall, F1 score, AUROC, and AUPRC, to reflect both classification quality and practical utility in threat detection settings. We conclude few-shot prompting improves performance across multiple LLMs.

\end{abstract}


\section{Introduction}
\label{sec:introduction}

Phishing remains a prevalent and evolving cybersecurity threat, with malicious URLs serving as a primary tool to deceive users into visiting fraudulent websites. These attacks often lead to credential theft, financial fraud, or data breaches. Traditional detection systems, particularly blacklist-based methods, fall short when faced with newly generated or obfuscated phishing URLs, as they are highly dependent on prior knowledge \cite{tian2025past}. Recent threat intelligence reports highlight the growing scale and sophistication of phishing attacks across industries~\cite{ibm2024xforce,hoxhunt2024trends}. This limitation has fueled the need for learning-based approaches that generalize better to unseen or adversarial examples.

Early research on phishing URL detection leveraged machine learning (ML) and deep learning (DL) by extracting the lexical and structural characteristics of URLs. These approaches commonly use classifiers such as decision trees, support vector machines, and neural networks \cite{omari2025advanced, kocyigit2024enhanced, ghalechyan2024phishing}. With the increasing complexity of phishing tactics, newer models have adopted deep neural architectures such as convolutional and recurrent networks that capture sequential patterns directly from raw URLs without manual feature engineering \cite{zara2024phishing, barik2025web}. Several works also employ optimization strategies like genetic algorithms and ensemble models to handle class imbalance and enhance model robustness \cite{remya2024effective, rafsanjani2024enhancing}.

Recently, LLMs have gained traction for cybersecurity tasks, including malicious URL detection. Pre-trained transformers such as BERT\cite{devlin2019bert} and GPT\cite{radford2018improving} can be adapted to URL-based tasks through fine-tuning or prompt-based classification \cite{mahdaouy2024domurls_bert, liu2025pmanet}. These models can capture contextual cues from input text, even when the data is minimal or unstructured. PhishURLDetect \cite{ali2025phishurldetect}, for example, shows that fine-tuning LLMs using parameter-efficient methods such as LoRA\cite{hu2022lora} can produce competitive performance while significantly reducing computational overhead.

Despite these advances, comparative evaluation of proprietary LLMs under standardized zero-shot and few-shot prompting settings remains underexplored. Prior studies either benchmark a single model or evaluate performance under varied conditions without consistent metrics or datasets \cite{nasution2025benchmarking}. Moreover, while real-world phishing data is often imbalanced, balanced datasets are commonly used in evaluation to enable fair assessment of model generalization and efficiency. Some recent models designed for malicious URL classification \cite{zhou2025integrated, senanayake2025madonna} and phishing webpage detection \cite{lee2024multimodal} highlight the potential of neural and transformer-based systems, but few investigate prompt-only inference in an instruction-tuned context.

In this work, we fill this gap by evaluating a range of proprietary LLMs on a balanced phishing URL dataset using prompt-based classification. Our experiments cover both zero-shot and few-shot scenarios\cite{brown2020language} across models such as GPT-4o (OpenAI), Claude-3-7-sonnet-20250219 (Anthropic), Grok-3-Beta(xAI). We report key evaluation metrics including accuracy, precision, recall, F1 score, AUROC, and AUPRC to assess model effectiveness and practical applicability in real-world phishing detection scenarios. In summary, our contributions to the community are as follows:

\begin{enumerate}
    \item  We present a unified benchmark of instruction-tuned proprietary LLMs for phishing URL detection under standardized zero-shot and few-shot prompting settings.
    \item We evaluate all models on a publicly available phishing URL dataset, which we balanced by sampling equal numbers of phishing and legitimate URLs, using a consistent and model-agnostic prompt design.
    \item We offer a comparative analysis of model performance and generalizability, and release our codebase to promote reproducibility and support further research in the community.
\end{enumerate}


\section{Related Work}

Phishing URL detection has been widely investigated across a variety of ML paradigms. Early approaches primarily relied on handcrafted lexical and host-based features to build classifiers such as decision trees, support vector machines, or ensemble methods \cite{omari2025advanced, kocyigit2024enhanced}. These models often struggled with generalization, especially in the presence of evolving phishing tactics and class imbalance. To address this, optimization strategies such as genetic algorithms, synthetic resampling (e.g., SMOTETomek\cite{omari2025advanced}), and feature weighting schemes have been proposed to enhance detection performance \cite{remya2024effective, rafsanjani2024enhancing}.

Deep learning-based methods have introduced greater flexibility by automatically learning hierarchical representations from raw URLs. Studies have applied convolutional and recurrent neural networks to extract local and sequential patterns \cite{zara2024phishing, barik2025web}, while recent designs incorporate multilayer perceptrons and attention mechanisms \cite{ghalechyan2024phishing}. Several efforts have focused on enhancing feature selection for neural architectures through empirical analysis and comparative evaluations \cite{ghalechyan2024phishing, kocyigit2024enhanced}. Despite their promise, many of these models are limited to static supervised training and are typically benchmarked on datasets that lack zero-shot capabilities.

LLMs have recently been explored for phishing detection, leveraging their ability to capture semantic cues in textual inputs. DomURLs\_BERT proposed a fine-tuned BERT-based classifier to detect both domains and URLs \cite{mahdaouy2024domurls_bert}, while PMANet introduced a post-trained transformer-based attention framework that incorporates multilevel semantic and lexical features \cite{liu2025pmanet}. These works show the adaptability of language models to security tasks, although they primarily focus on traditional fine-tuning, and do not evaluate zero- or few-shot settings.

The parameter-efficient PhishURLDetect framework applied LoRA to adapt large models to phishing detection with reduced computational cost, but the evaluation was limited to a single instruction-tuned model and did not compare against LLMs \cite{ali2025phishurldetect}. Meanwhile, MADONNA employed neural networks with browser-level telemetry to detect malicious domains in real-time but was not designed to operate in prompt-based configurations \cite{senanayake2025madonna}.

A broader benchmarking effort by \cite{nasution2025benchmarking} evaluated twenty-one open-source LLMs using prompt engineering for phishing link detection. However, the study excluded commercial models and was restricted to English-language prompts. Similarly, \cite{zhou2025integrated} proposed a CSPPC-BiLSTM framework that fused handcrafted features with sequence modeling but did not address inference efficiency or few-shot generalization. Few works have investigated prompt-only inference for phishing detection, although \cite{lee2024multimodal} introduced a multimodal vision-language framework for phishing webpage identification.

Finally, the comprehensive survey by \cite{tian2025past} provides an overview of malicious URL detection, available datasets, and public codebases. Highlights the lack of unified benchmarking among emerging LLM-based methods and underscores the need to evaluate models under consistent experimental setups. This motivates the need for our work, which systematically compares zero-shot and few-shot prompting across a representative set of LLMs on a balanced dataset of phishing and legitimate URLs.

\section{Methodology}
\label{sec:methodology}

\begin{figure*}[h]
    \centering
    \begin{tikzpicture}[
        node distance = 1.1cm,
        box/.style={
            rectangle,
            draw=black!80, 
            line width=1.2pt,
            text centered, 
            minimum height=1.0cm, 
            font=\large,
            inner sep=4pt
        },
        urlbox/.style={box, fill=orange!25, text width=1.5cm},
        decision/.style={
            diamond, 
            draw=black!80, 
            line width=1.2pt,
            aspect=1.4, 
            fill=pink!25, 
            text width=1.5cm, 
            text centered,
            font=\large
        },
        promptbox/.style={box, fill=yellow!20, text width=2.6cm, minimum height=1.5cm, font=\normalsize},
        llmbox/.style={box, fill=purple!15, text width=1.5cm, rounded corners=3pt, minimum height=1.4cm},
        procbox/.style={box, fill=blue!20, text width=1.5cm, rounded corners=3pt},
        outbox/.style={box, fill=green!25, text width=1.9cm},
        metricsbox/.style={box, fill=green!25, text width=2.2cm, minimum height=0.9cm},
        arrow/.style={->, >=Stealth, line width=1.2pt, draw=black!80},
        label/.style={font=\normalsize, text=black}
    ]

    \node[urlbox] (url) at (0,2) {$u$};
    \node[decision, right=0.95cm of url] (mode) {Mode};
    
    \node[promptbox, above right=0.4cm and 0.7cm of mode] (zs) {
        $\mathcal{P}_\text{ZS}(u) =$\\[4pt]
        $\mathcal{I} \oplus \mathcal{Q}(u)$
    };
    
    \node[promptbox, below right=0.4cm and 0.7cm of mode] (fs) {
        $\mathcal{P}_\text{FS}(u) =$\\[4pt]
        $\mathcal{I} \oplus \mathcal{E} \oplus \mathcal{Q}(u)$
    };
    
    \coordinate (merge_x) at ($(zs.east)!0.5!(fs.east) + (1.25cm,0)$);
    \node[llmbox] (llm) at (merge_x |- mode) {
        $\mathcal{M}$
    };
    
    \node[procbox, right=1.4cm of llm] (parse) {
        $\hat{y}$
    };
    
    \node[outbox, below=1.3cm of parse] (pred) {
        $\hat{y} \in \{0,1\}$
    };
    
    \node[metricsbox, below=1.0cm of pred] (metrics) {
        Evaluate
    };
    
    \draw[arrow] (url) -- (mode);
    
    \draw[arrow] (mode) |- node[pos=0.2, left, label, xshift=-2pt] {ZS} (zs);
    
    \draw[arrow] (mode) |- node[pos=0.2, left, label, xshift=-2pt] {FS} (fs);
    
    \draw[arrow] (zs.east) -| (llm.north);
    \draw[arrow] (fs.east) -| (llm.south);
    
    \draw[arrow] (llm) -- (parse);
    
    \draw[arrow] (parse) -- (pred);
    
    \draw[arrow] (pred) -- (metrics);
    
    \draw[arrow, dashed, line width=1.0pt] (metrics.south) 
        -- ++(0,-0.7)
        -| node[pos=0.2, below, label] {repeat} 
        ($(url.south) + (0,-0.3)$)
        -- (url.south);

    \end{tikzpicture}
    \caption{Phishing URL classification methodology. Each URL $u$ is processed via zero-shot ($\mathcal{P}_\text{ZS}$) or few-shot ($\mathcal{P}_\text{FS}$) prompting. Zero-shot prompts contain task instruction $\mathcal{I}$ and query $\mathcal{Q}(u)$; few-shot prompts additionally include examples $\mathcal{E}$. The LLM $\mathcal{M}$ generates responses parsed to binary predictions $\hat{y} \in \{0,1\}$ (0=phishing, 1=legitimate). Predictions are evaluated against ground truth $y_i$ using macro-averaged accuracy, precision, recall, F1-score, AUROC, and AUPRC across all test URLs.}
    \label{fig:methodology}
\end{figure*}
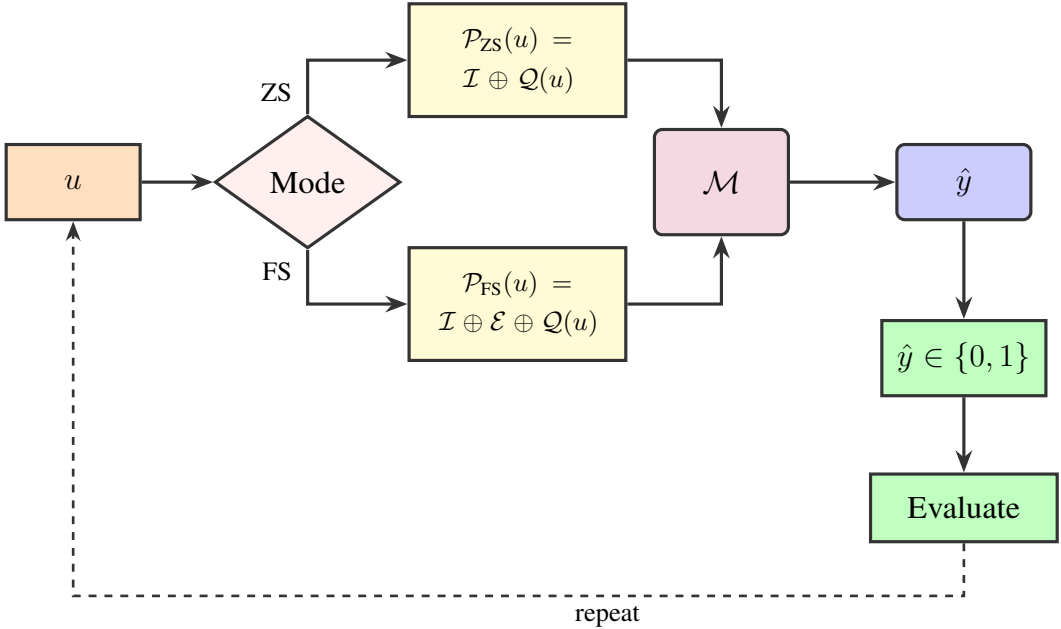

We evaluate three large language models (GPT-4o, Claude-3.7-sonnet-20250219, and Grok-3-Beta) on phishing URL classification under zero-shot and few-shot prompting. Experiments are conducted on both balanced and imbalanced datasets to assess performance across different class distributions. Figure~\ref{fig:methodology} illustrates the complete pipeline.

\subsection{Experimental Setup}
\label{sec:experimental_setup}

We use the PhiUSIIL Phishing URL dataset~\cite{phiusiil_phishing_url_(website)_967}, preprocessing to retain URL strings and binary labels (0=phishing, 1=legitimate). We conduct two sets of experiments to evaluate model performance under different class distributions.

For balanced evaluation, we construct a corpus of 10,000 URLs by randomly sampling 5,000 phishing and 5,000 legitimate URLs using seed 42. In zero-shot experiments, all 10,000 samples are used for evaluation. In few-shot experiments, we sample 6 examples (3 phishing, 3 legitimate) as $\mathcal{E}$ and evaluate on the remaining 9,994 samples, ensuring examples are disjoint from the evaluation set.

For imbalanced evaluation, we construct test sets of 1,000 URLs with phishing ratios of 1\% and 10\%. Each ratio is tested with two random seeds (S123, S456) to verify stability across different samples. In zero-shot experiments, all 1,000 samples are used for evaluation. In few-shot experiments under 10\% imbalance with seed S123, we vary the number of examples in $\mathcal{E}$ using 1, 3, or 9 examples and evaluate on the remaining samples.


\subsection{Prompt Construction}
\label{sec:prompt_construction}

Each prompt consists of three components: a task instruction $\mathcal{I}$, a query $\mathcal{Q}(u)$ for the target URL, and optionally, a set of examples $\mathcal{E}$ for few-shot learning. The task instruction is:
\begin{quote}
\small
You are a cybersecurity expert. Respond only with 0 for phishing or 1 for legitimate.
\end{quote}

The query for each URL $u$ is:
\begin{quote}
\small
URL: \texttt{\{u\}}\\
Is this URL phishing or legitimate? Respond with 0 or 1.
\end{quote}

For zero-shot prompting, we construct $\mathcal{P}_\text{ZS}(u) = \mathcal{I} \oplus \mathcal{Q}(u)$, where $\oplus$ denotes concatenation. For few-shot prompting, we construct $\mathcal{P}_\text{FS}(u) = \mathcal{I} \oplus \mathcal{E} \oplus \mathcal{Q}(u)$, where each example in $\mathcal{E}$ is formatted as:
\begin{quote}
\small
URL: \texttt{\{u'\}}\\
Answer: \texttt{\{y'\}} (\textit{label})
\end{quote}
with $y' \in \{0,1\}$ and label text "phishing" for $y'=0$ or "legitimate" for $y'=1$.

For GPT-4o and Grok-3-Beta, zero-shot prompts consist of a system message containing the task instruction, followed by a user message with the URL query. Few-shot prompts use the same system message, followed by one user message per example, then a final user message with the target URL query. For Claude-3.7-sonnet, zero-shot prompts concatenate the task instruction and query separated by double newlines. Few-shot prompts concatenate the instruction, each example (also separated by double newlines), then the query.


\subsection{Evaluation Protocol}
\label{sec:evaluation_protocol}

We access the three models via their official APIs: GPT-4o (OpenAI), Claude-3.7-sonnet-20250219 (Anthropic), and Grok-3-Beta (xAI). All inference is performed with temperature set to 0 and maximum output tokens limited to 10. Model responses are parsed to extract binary predictions $\hat{y} \in \{0,1\}$. Responses that cannot be parsed are excluded from evaluation.

We compute six metrics: accuracy, macro-averaged precision, recall, F1-score, AUROC, and AUPRC. Macro-averaging computes per-class metrics and averages them, treating both classes equally. Let $\mathcal{C} = \{0,1\}$ denote the class set. For class $c$, true positives ($\text{TP}_c$), false positives ($\text{FP}_c$), and false negatives ($\text{FN}_c$) are used to compute precision $\text{TP}_c / (\text{TP}_c + \text{FP}_c)$, recall $\text{TP}_c / (\text{TP}_c + \text{FN}_c)$, and F1-score $2 \cdot \text{Precision}_c \cdot \text{Recall}_c / (\text{Precision}_c + \text{Recall}_c)$. Macro-averaged metrics average these values across both classes. All metrics are implemented using scikit-learn~\cite{pedregosa2011scikit}.


\section{Experimental Results}

We evaluate three large language models (GPT-4o, Claude-3.7-sonnet-20250219, and Grok-3-Beta) on phishing URL classification under zero-shot and few-shot settings. Experiments are conducted on both balanced and imbalanced datasets to assess model performance across different class distributions. We report Accuracy, Precision, Recall, and F1 Score, with all metrics macro-averaged unless otherwise stated.

\subsection{Balanced Dataset Evaluation}
\label{sec:balanced}

On a balanced test set of 10,000 URLs, few-shot prompting with six examples substantially improves performance across all models (Table~\ref{tab:results_balanced}). Grok-3-Beta achieves the strongest few-shot performance, leading in five of six metrics: Accuracy (0.9405), Precision (0.9492), F1 (0.9399), AUROC (0.9405), and AUPRC (0.9573). Claude-3.7 attains the highest Recall (0.9526) but lower Precision (0.9027) compared to Grok-3-Beta. GPT-4o demonstrates consistent gains across metrics but trails the other models in overall performance.

Grok-3-Beta exhibits a notable precision-recall trade-off when transitioning from zero-shot to few-shot settings: Recall decreases from 0.9735 to 0.9307 while Precision increases from 0.8361 to 0.9492, indicating stricter classification thresholds in few-shot mode.

\begin{table}[H]
\centering
\caption{Performance comparison on balanced phishing URL detection (10,000 URLs). All metrics are macro-averaged.}
\label{tab:results_balanced}
\begin{tabular}{@{}llcccccc@{}}
\toprule
\textbf{Model} & \textbf{Setting} & \textbf{Accuracy} & \textbf{Precision} & \textbf{Recall} & \textbf{F1} & \textbf{AUROC} & \textbf{AUPRC} \\
\midrule
GPT-4o & Zero-shot  & 0.8752 & 0.8421 & 0.9232 & 0.8808 & 0.8752 & 0.9018 \\
GPT-4o & Few-shot   & 0.9050 & 0.8880 & 0.9270 & 0.9071 & 0.9050 & 0.9258 \\
\addlinespace
Claude-3.7 & Zero-shot & 0.8759 & 0.8778 & 0.8734 & 0.8756 & 0.8759 & 0.9072 \\
Claude-3.7 & Few-shot  & 0.9250 & 0.9027 & 0.9526 & 0.9270 & 0.9250 & 0.9395 \\
\addlinespace
Grok-3-Beta & Zero-shot & 0.8914 & 0.8361 & \colorbox{gray!20}{\,0.9735\,} & 0.8996 & 0.8914 & 0.9114 \\
Grok-3-Beta & Few-shot  & \colorbox{gray!20}{\,0.9405\,} & \colorbox{gray!20}{\,0.9492\,} & 0.9307 & \colorbox{gray!20}{\,0.9399\,} & \colorbox{gray!20}{\,0.9405\,} & \colorbox{gray!20}{\,0.9573\,} \\
\bottomrule
\end{tabular}
\end{table}
\FloatBarrier


\subsubsection{Per-Class Evaluation}
\label{sec:perclass}


\begin{figure*}[!htb]  
\centering
\subfloat[GPT-4o: Zero-shot]{\includegraphics[width=0.30\linewidth]{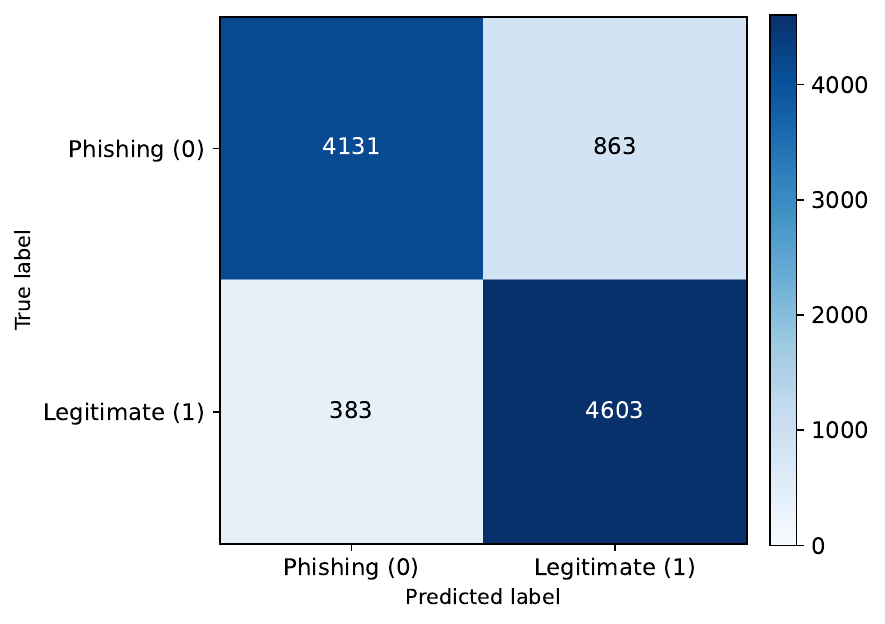}}
\hspace{3mm}
\subfloat[GPT-4o: Few-shot]{\includegraphics[width=0.30\linewidth]{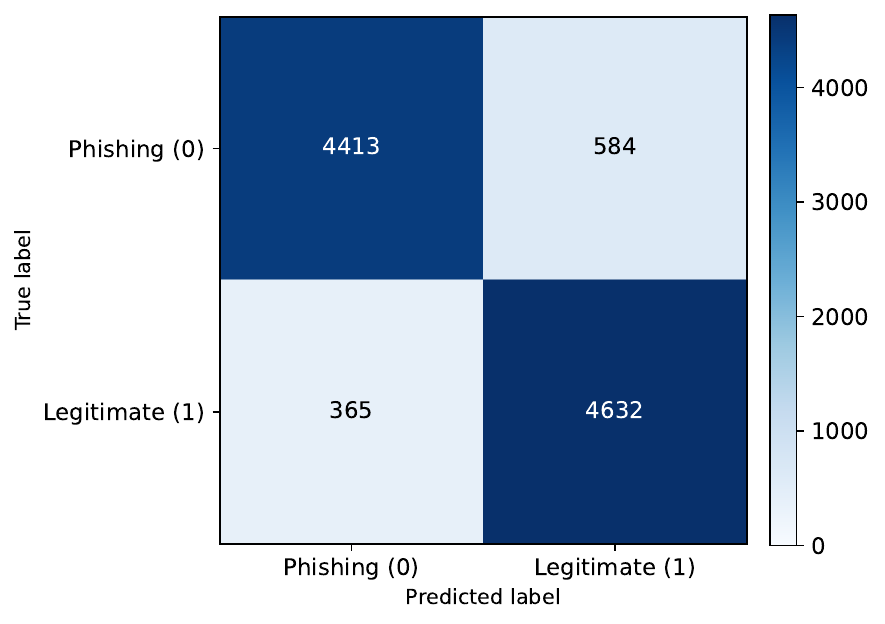}}
\hspace{3mm}
\subfloat[Claude: Zero-shot]{\includegraphics[width=0.30\linewidth]{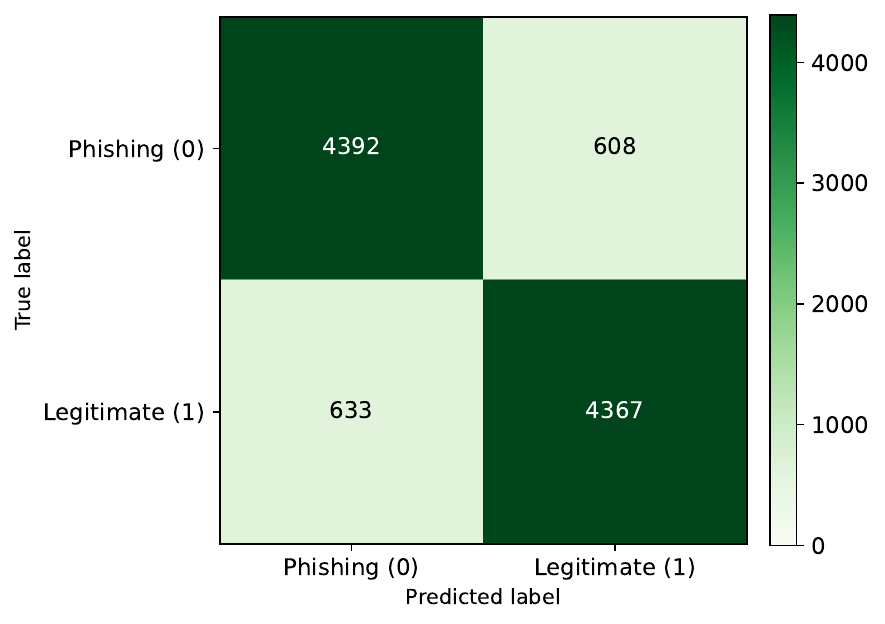}}

\vspace{5mm}

\subfloat[Claude: Few-shot]{\includegraphics[width=0.30\linewidth]{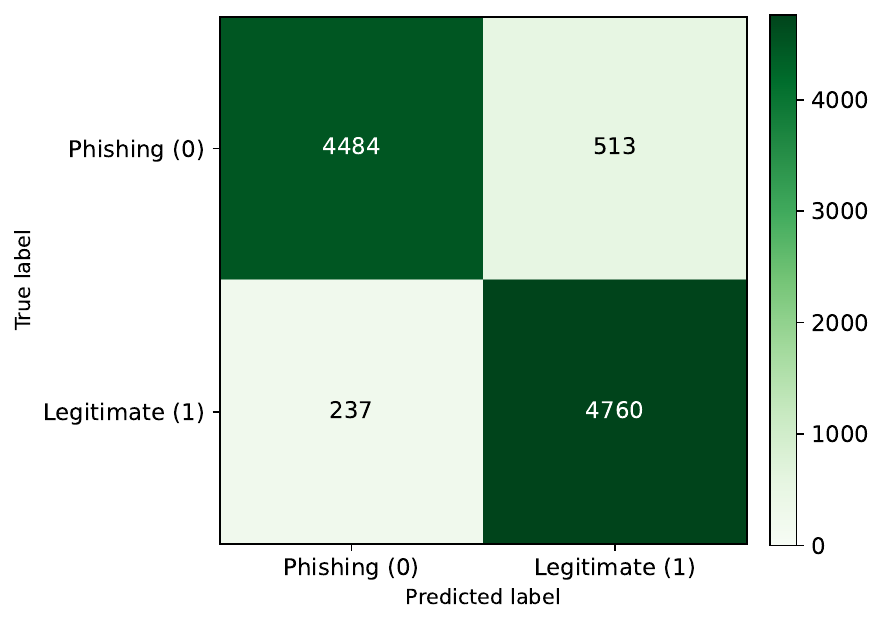}}
\hspace{3mm}
\subfloat[Grok-3: Zero-shot]{\includegraphics[width=0.30\linewidth]{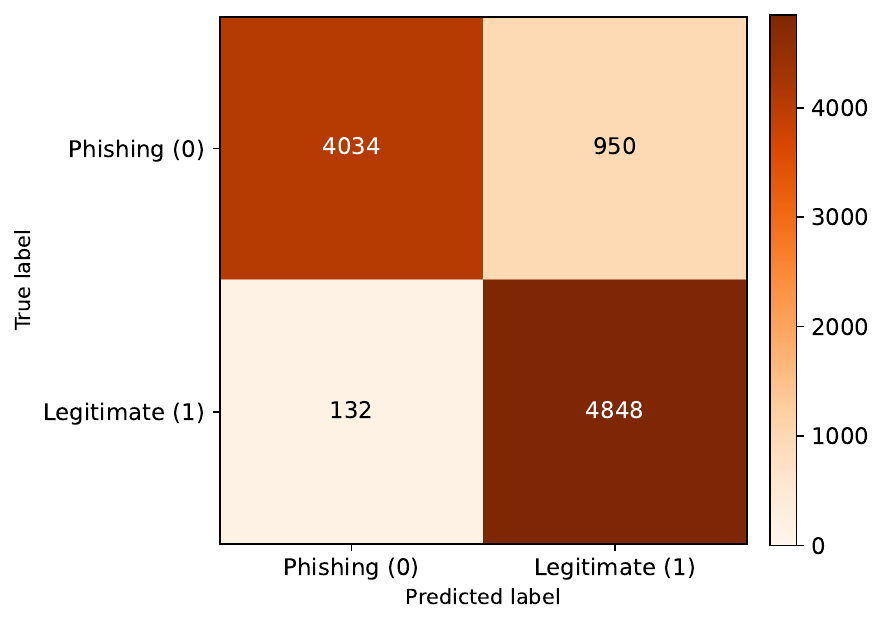}}
\hspace{3mm}
\subfloat[Grok-3: Few-shot]{\includegraphics[width=0.30\linewidth]{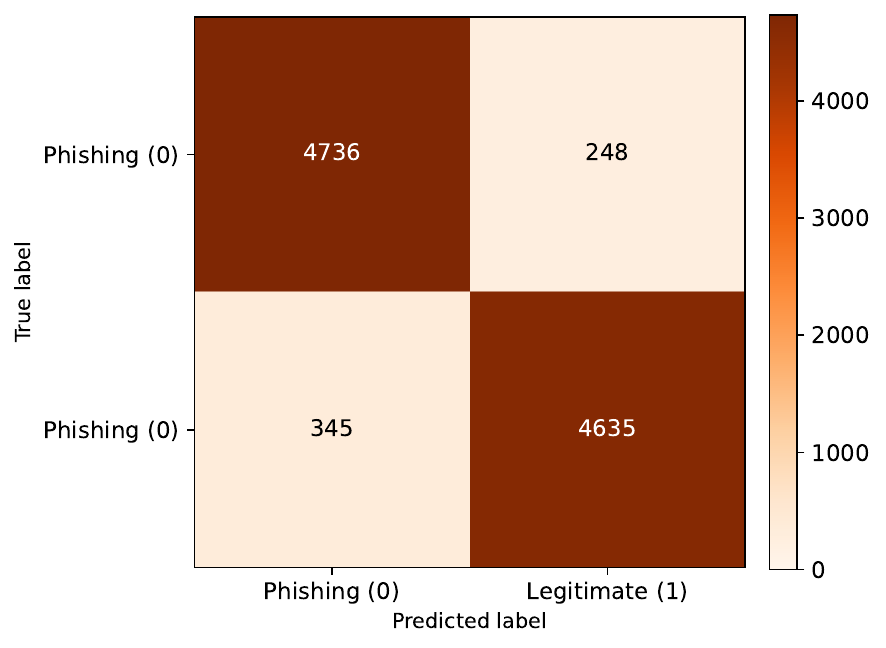}}

\vspace{1mm}
\caption{Confusion matrices for all models under zero-shot and few-shot prompting.}
\label{fig:all_conf_matrices}
\end{figure*}
\FloatBarrier

Figure~\ref{fig:all_conf_matrices} shows confusion matrices for all models. Few-shot prompting with six examples reduces false negatives across all models. Grok-3-Beta demonstrates the largest improvement, with false negatives dropping from 950 to 248. GPT-4o and Claude-3.7 reduce false negatives from 863 to 584 and 608 to 513, respectively. Grok-3-Beta few-shot achieves the fewest false negatives (248), while Grok-3-Beta zero-shot achieves the fewest false positives (132). Claude-3.7 few-shot reduces false positives to 237. These results reflect the precision-recall trade-off in Table~\ref{tab:results_balanced}.


\subsubsection{AUROC and AUPRC Curves}
\label{sec:curves}

\begin{figure*}[h]
\centering
\includegraphics[width=0.23\linewidth]{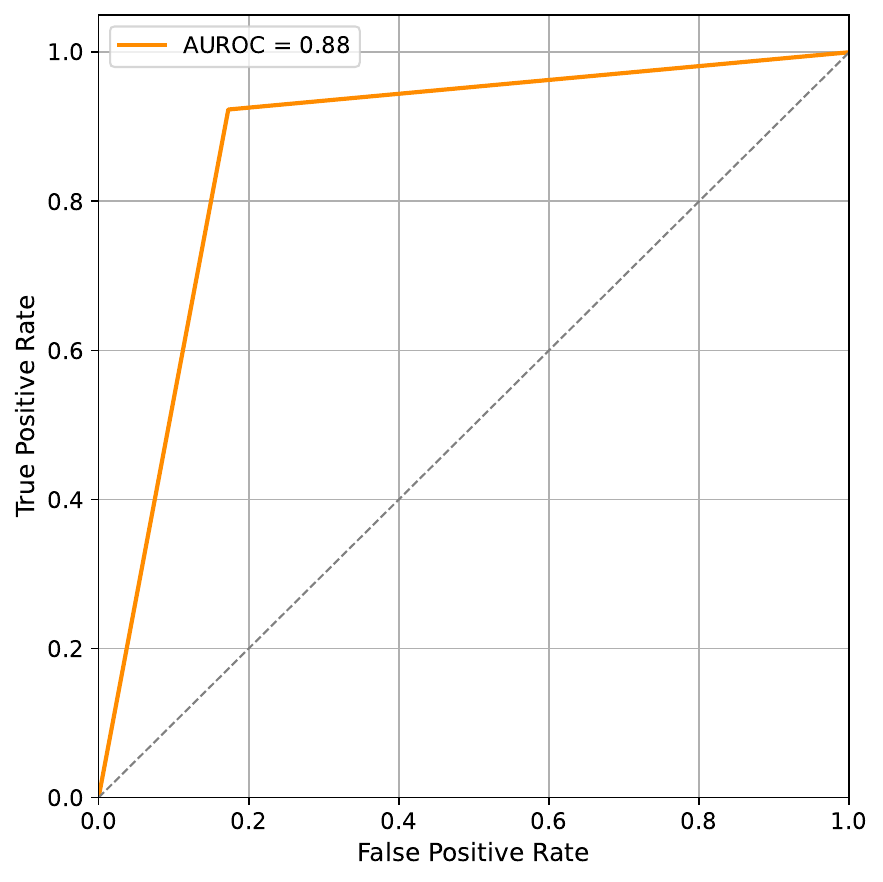}
\hspace{2mm}
\includegraphics[width=0.23\linewidth]{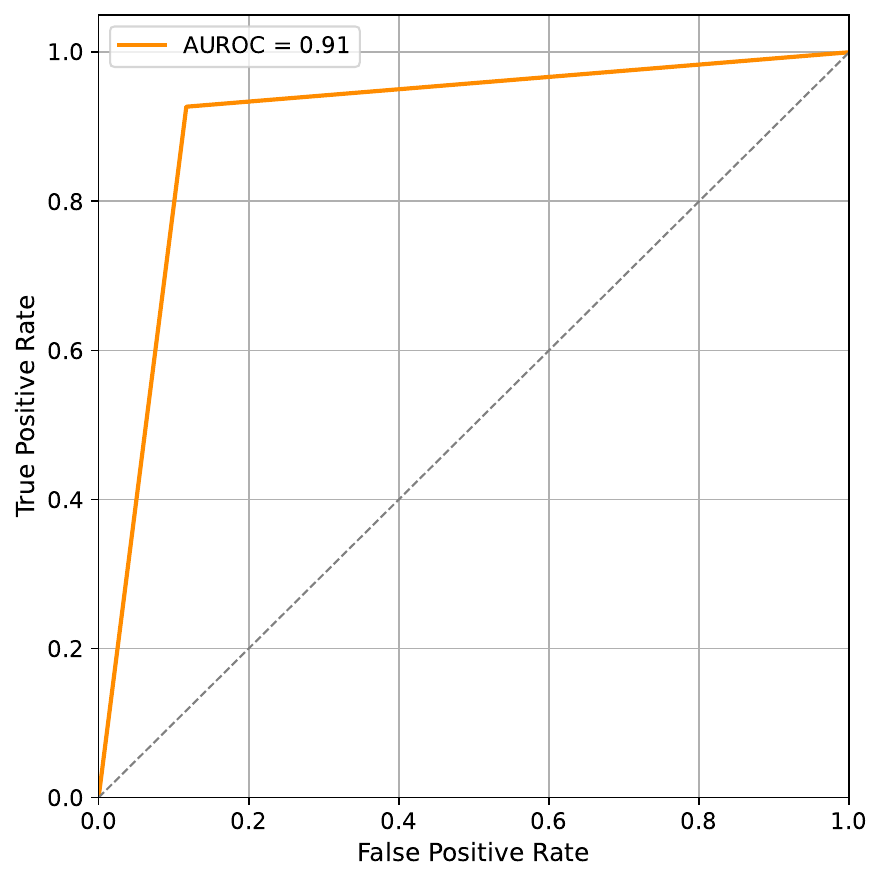}
\hspace{2mm}
\includegraphics[width=0.23\linewidth]{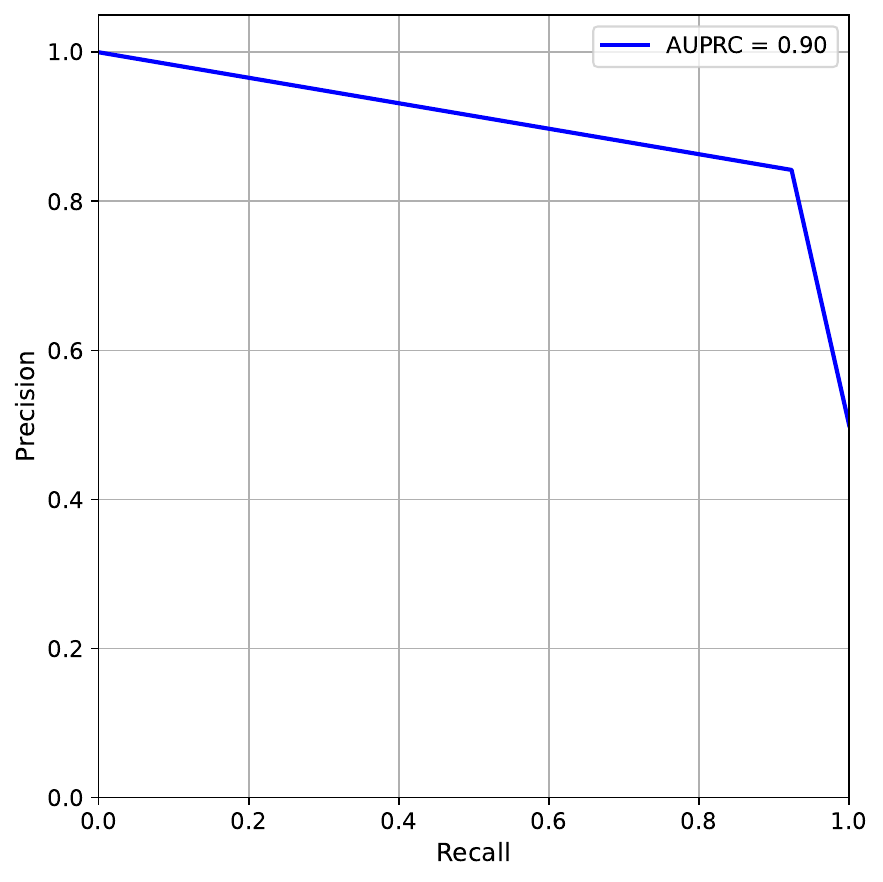}
\hspace{2mm}
\includegraphics[width=0.23\linewidth]{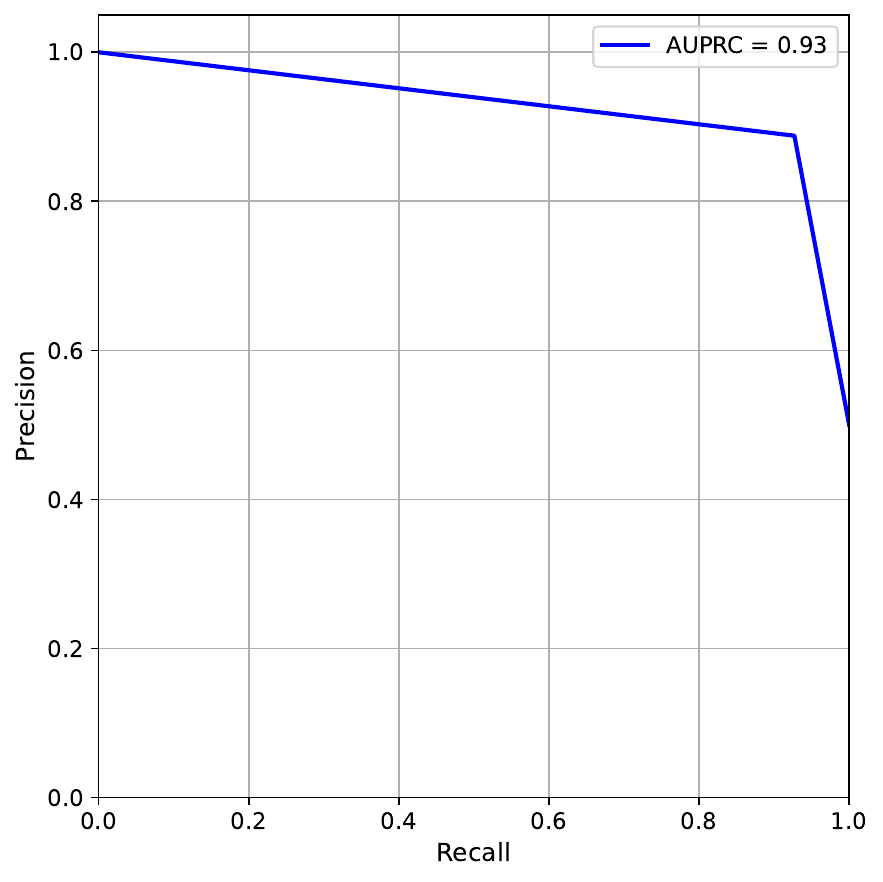}
\vspace{1mm}
\caption{GPT-4o ROC (left) and PR (right) curves under zero-shot and few-shot prompting.}
\label{fig:gpt4o_curves}
\end{figure*}

\begin{figure*}[!htbp]
\centering
\includegraphics[width=0.23\linewidth]{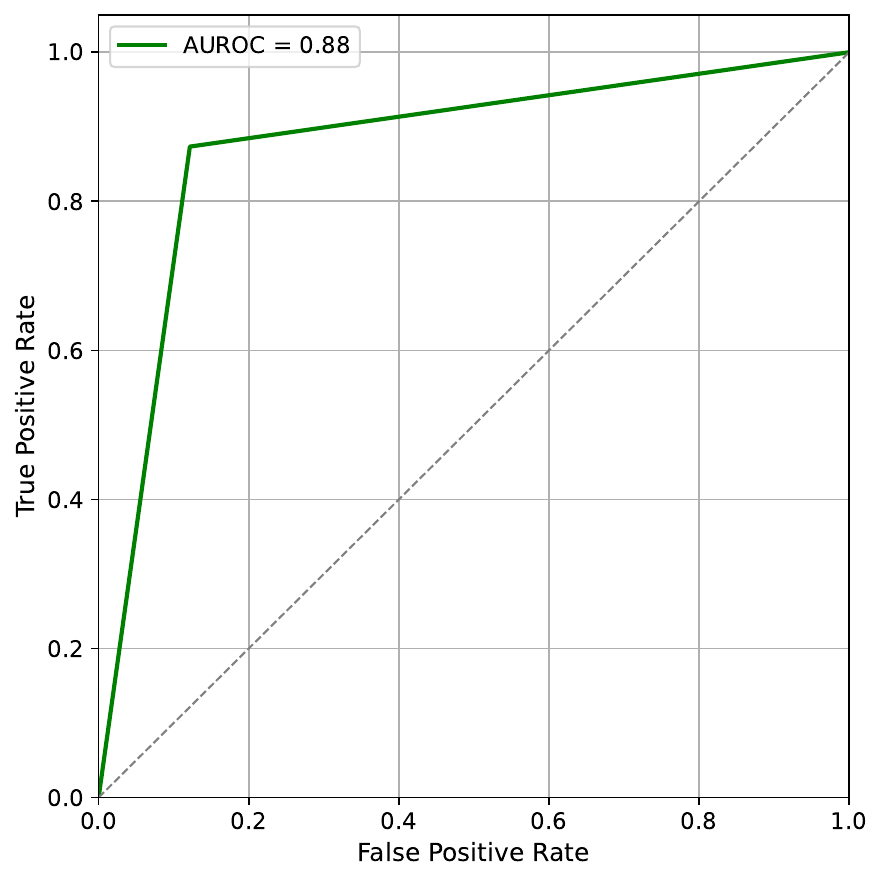}
\hspace{2mm}
\includegraphics[width=0.23\linewidth]{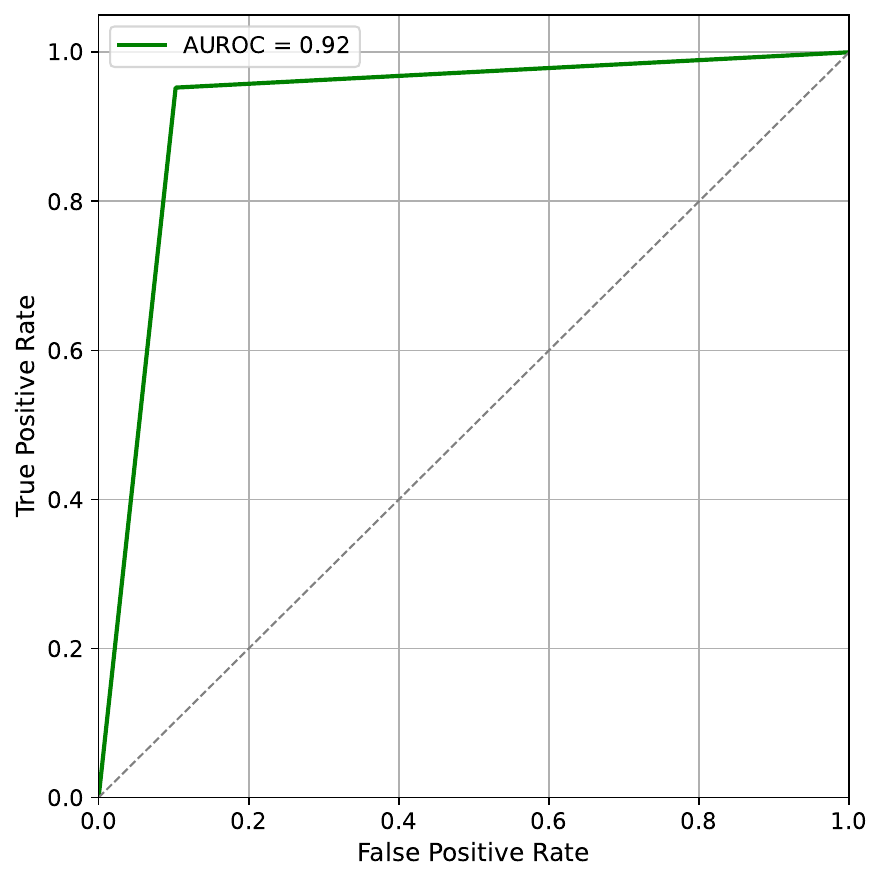}
\hspace{2mm}
\includegraphics[width=0.23\linewidth]{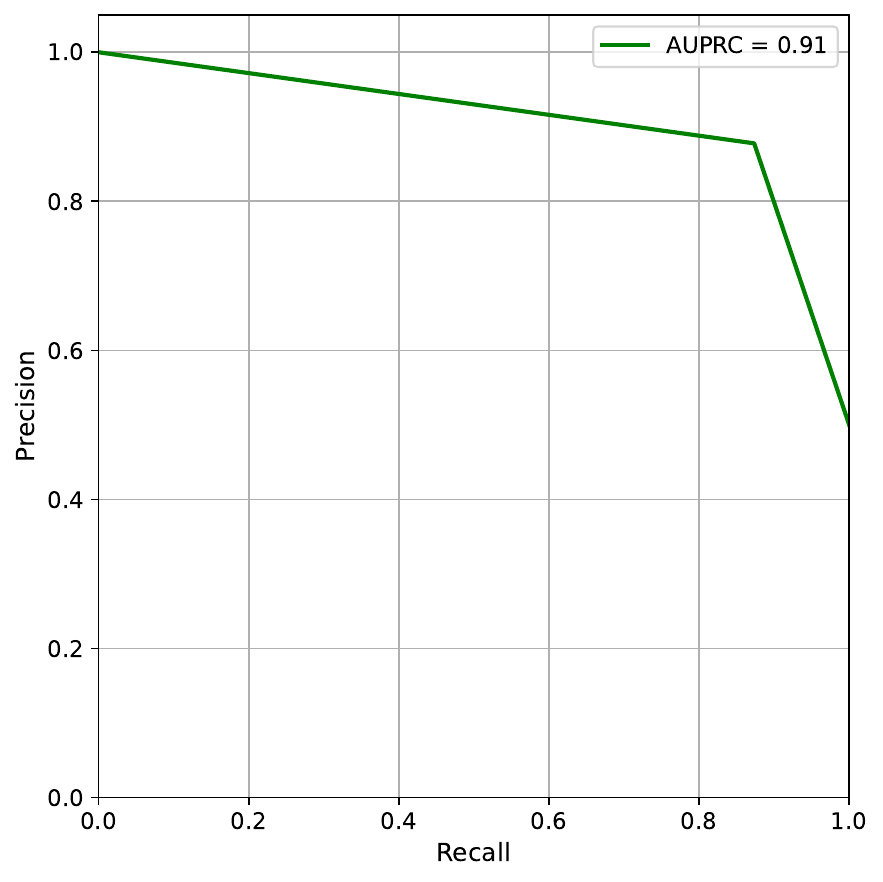}
\hspace{2mm}
\includegraphics[width=0.23\linewidth]{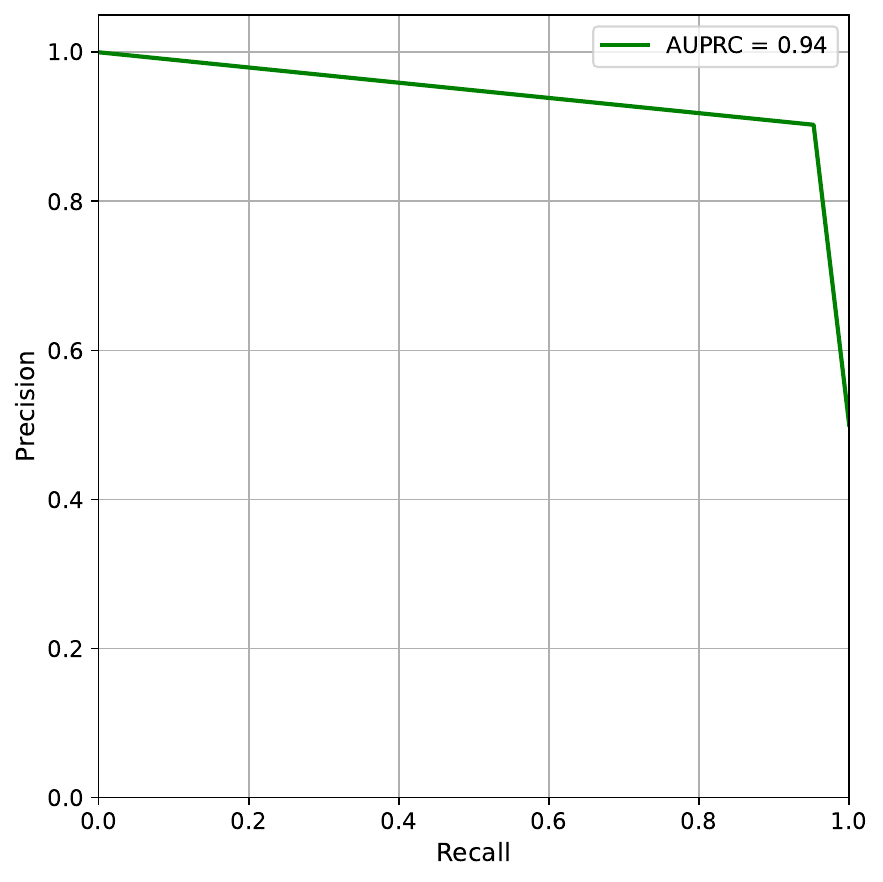}
\vspace{1mm}
\caption{Claude-3.7 ROC (left) and PR (right) curves under zero-shot and few-shot prompting.}
\label{fig:claude_curves}
\end{figure*}

\begin{figure*}[!htbp]
\centering
\includegraphics[width=0.23\linewidth]{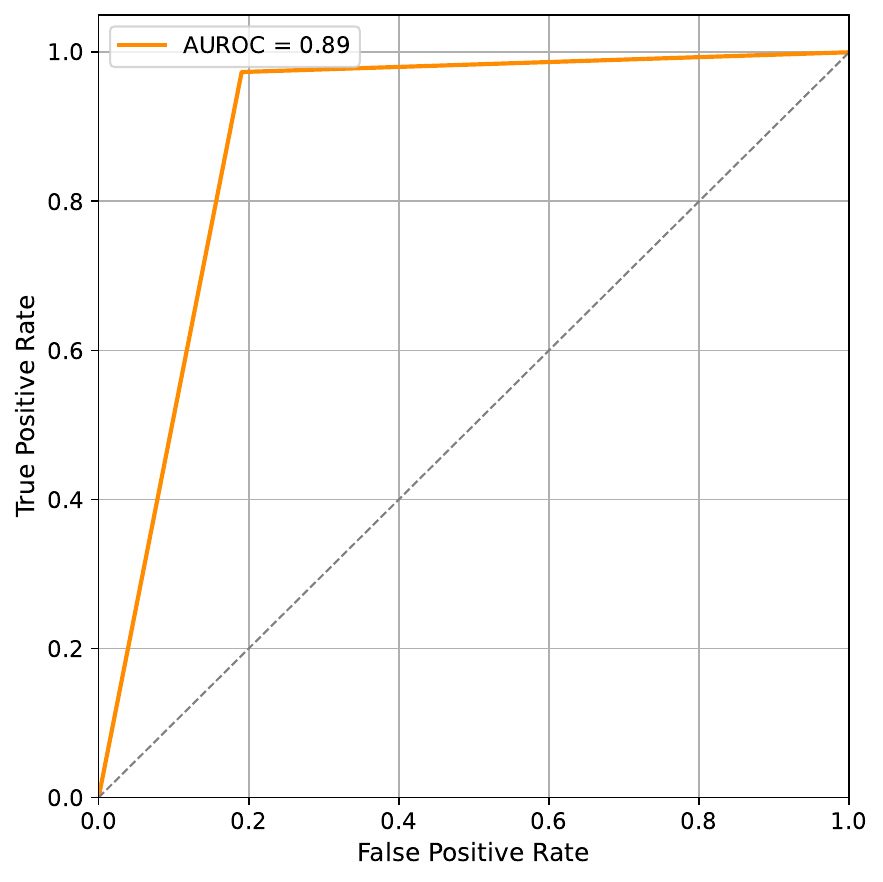}
\hspace{2mm}
\includegraphics[width=0.23\linewidth]{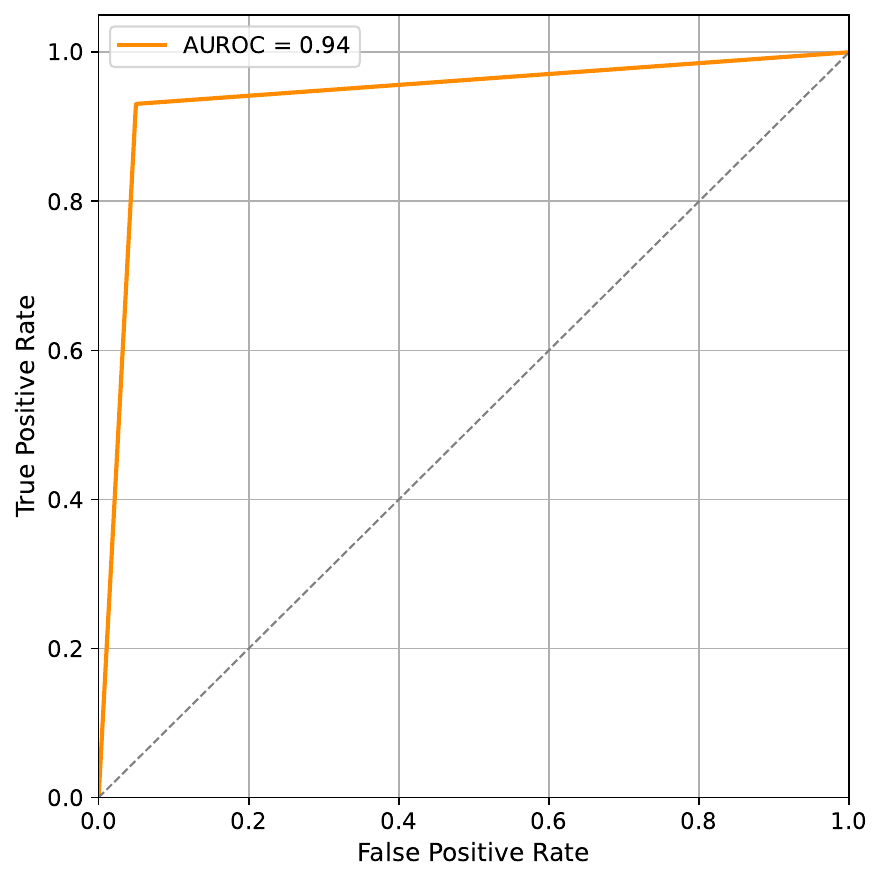}
\hspace{2mm}
\includegraphics[width=0.23\linewidth]{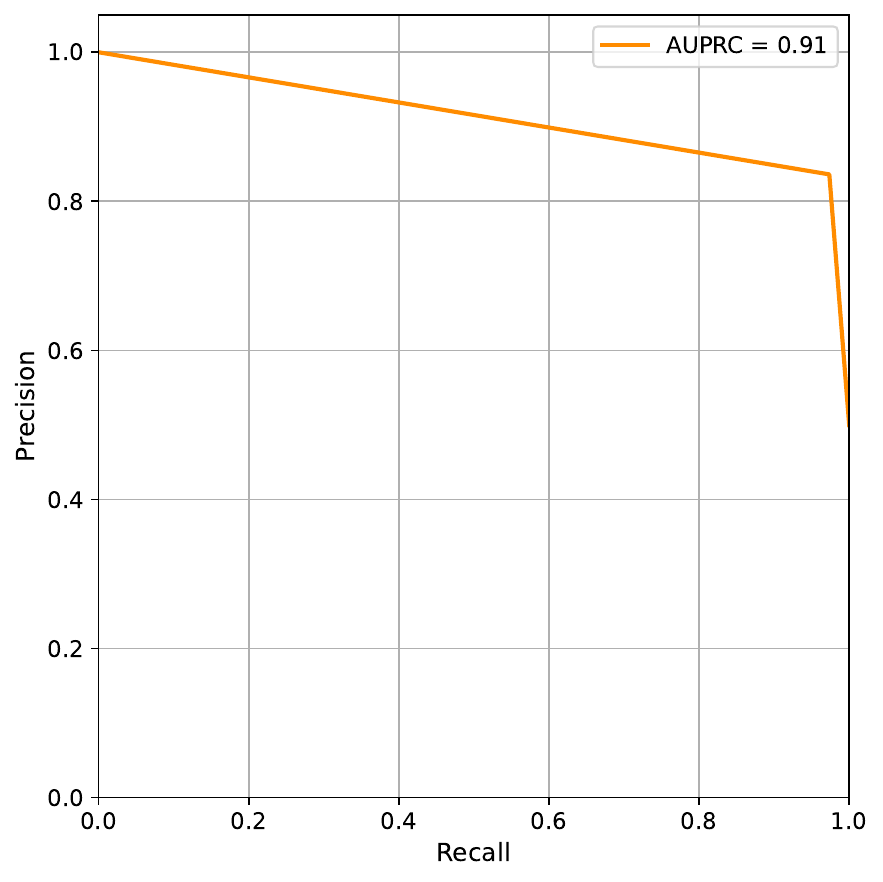}
\hspace{2mm}
\includegraphics[width=0.23\linewidth]{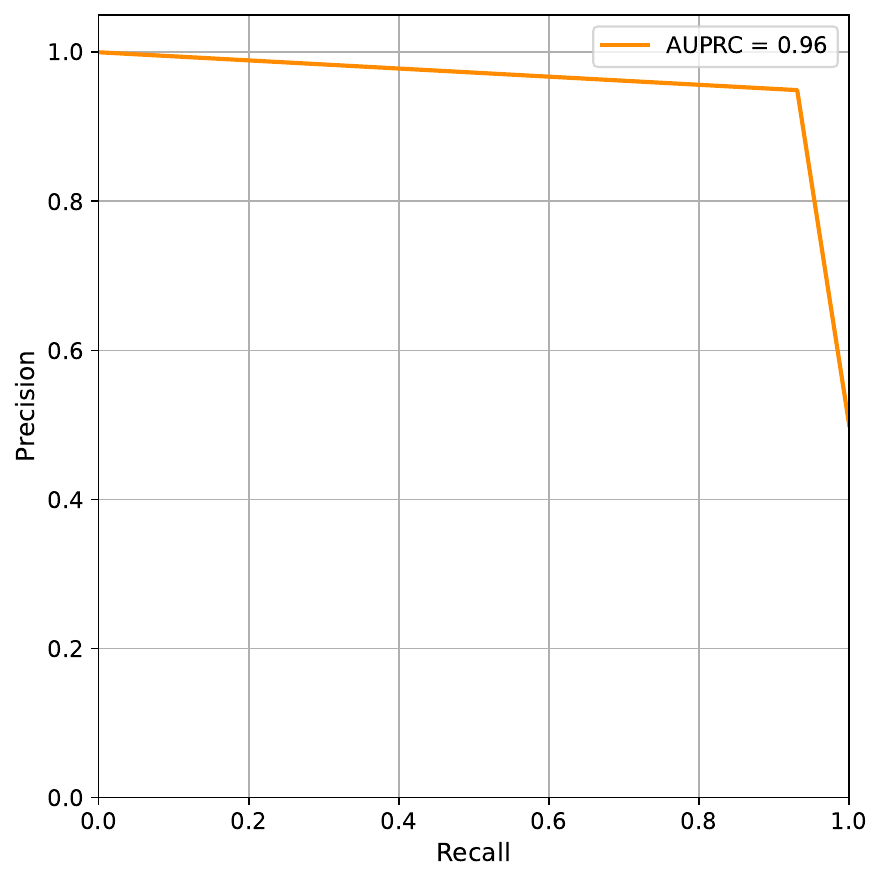}
\vspace{1mm}
\caption{Grok-3-Beta ROC (left) and PR (right) curves under zero-shot and few-shot prompting.}
\label{fig:grok3_curves}
\end{figure*}

ROC and Precision-Recall curves for all models under zero-shot and few-shot prompting are presented in Figures~\ref{fig:gpt4o_curves} to~\ref{fig:grok3_curves}. Few-shot prompting with six examples consistently improves AUROC and AUPRC across models.

Grok-3-Beta achieves the highest few-shot performance (AUROC: 0.9405, AUPRC: 0.9573) with a steep ROC rise and stable PR curve, indicating strong discriminative ability. Claude-3.7 shows substantial improvements from zero-shot (AUROC: 0.8759, AUPRC: 0.9072) to few-shot (AUROC: 0.9250, AUPRC: 0.9395), demonstrating effective learning from examples. GPT-4o exhibits consistent gains, with AUROC improving from 0.8752 to 0.9050 and AUPRC from 0.9018 to 0.9258. These curves illustrate model behavior across classification thresholds, complementing the scalar metrics presented earlier.

\FloatBarrier

\FloatBarrier

\FloatBarrier


\subsection{Imbalanced Dataset Evaluation}
\label{sec:imbalanced}

Increasing phishing URL proportion from 1\% to 10\% improves F1 scores across all models (Table~\ref{tab:results_imbalanced}). GPT-4o improves from 0.559 to 0.785, Claude-3.7 from 0.534 to 0.761, and Grok-3-Beta from 0.657 to 0.854. Models demonstrate robust performance across random seeds S123 and S456, with F1 differences below 0.05 at 10\% imbalance.

Few-shot learning with varying numbers of examples (1, 3, or 9) exhibits distinct patterns per model. Grok-3-Beta peaks at 1 example (F1: 0.906, Precision: 0.949) then degrades with 3 examples (F1: 0.821) and 9 examples (F1: 0.831). Claude-3.7 improves overall from 1 example (F1: 0.857) to 9 examples (F1: 0.876), with intermediate performance at 3 examples (F1: 0.842). GPT-4o gains consistently from 1 example (F1: 0.709) through 3 examples (F1: 0.801) to 9 examples (F1: 0.861). Grok-3-Beta dominates zero-shot settings with accuracy 0.976 and recall 0.938.

\begin{table}[h]
\centering
\caption{Performance comparison of LLMs on phishing URL detection across zero-shot and few-shot settings. All experiments use 1,000 test samples. All metrics except Accuracy are macro-averaged across phishing and legitimate classes. S123 and S456 denote random seeds 123 and 456, respectively. Imbalance ratios (1\% and 10\%) indicate the proportion of phishing URLs in the test set. Few-shot experiments use seed 123 with 10\% imbalance and vary the number of examples ($\mathcal{E}=1, 3, 9$).}
\label{tab:results_imbalanced}
\resizebox{\textwidth}{!}{%
\begin{tabular}{ll|cccc|ccc}
\toprule
\multirow{2}{*}{\textbf{Model}} & \multirow{2}{*}{\textbf{Metric}} & \multicolumn{4}{c|}{\textbf{Zero-Shot}} & \multicolumn{3}{c}{\textbf{Few-Shot}} \\
\cmidrule(lr){3-6} \cmidrule(lr){7-9}
& & \textbf{S123-1\%} & \textbf{S123-10\%} & \textbf{S456-1\%} & \textbf{S456-10\%} & \textbf{\(\mathcal{E}\)=1} & \textbf{\(\mathcal{E}\)=3} & \textbf{\(\mathcal{E}\)=9} \\
\midrule
\multirow{4}{*}{\textbf{GPT-4o}} 
& Accuracy  & 0.917 & 0.902 & 0.935 & 0.927 & 0.833 & 0.908 & 0.942 \\
& Precision & 0.544 & 0.742 & 0.561 & 0.789 & 0.676 & 0.754 & 0.821 \\
& Recall    & 0.859 & 0.866 & 0.918 & 0.888 & 0.863 & 0.891 & 0.919 \\
& F1-Score  & 0.559 & 0.785 & 0.591 & 0.828 & 0.709 & 0.801 & 0.861 \\
\midrule
\multirow{4}{*}{\textbf{Claude-3.7}} 
& Accuracy  & 0.881 & 0.879 & 0.903 & 0.903 & 0.945 & 0.933 & 0.951 \\
& Precision & 0.535 & 0.716 & 0.542 & 0.749 & 0.837 & 0.801 & 0.846 \\
& Recall    & 0.890 & 0.879 & 0.902 & 0.911 & 0.881 & 0.905 & 0.915 \\
& F1-Score  & 0.534 & 0.761 & 0.553 & 0.799 & 0.857 & 0.842 & 0.876 \\
\midrule
\multirow{4}{*}{\textbf{Grok-3-Beta}} 
& Accuracy  & 0.964 & 0.945 & \colorbox{gray!20}{\,0.976\,} & 0.962 & 0.969 & 0.915 & 0.924 \\
& Precision & 0.602 & 0.839 & 0.640 & 0.881 & \colorbox{gray!20}{\,0.949\,} & 0.768 & 0.783 \\
& Recall    & 0.932 & 0.872 & \colorbox{gray!20}{\,0.938\,} & 0.921 & 0.872 & 0.931 & 0.918 \\
& F1-Score  & 0.657 & 0.854 & 0.708 & 0.900 & \colorbox{gray!20}{\,0.906\,} & 0.821 & 0.831 \\
\bottomrule
\end{tabular}%
}
\end{table}
\FloatBarrier


\section{Conclusion}

Phishing attacks continue to evolve in sophistication, challenging traditional detection methods that depend on labeled datasets and manual feature engineering. Zero-shot and few-shot learning with large language models offer a practical alternative when labeled data is scarce or phishing tactics evolve rapidly. We benchmarked three large language models (GPT-4o, Claude-3.7-sonnet-20250219, and Grok-3-Beta) on phishing URL detection using prompt-based classification under both zero-shot and few-shot settings.

Our experiments on a balanced dataset of 10,000 URLs demonstrate that few-shot prompting with six examples substantially improves performance. Grok-3-Beta achieves the highest few-shot accuracy (0.9405) and F1 score (0.9399), while Claude-3.7-sonnet yields the best recall (0.9526). Under imbalanced conditions with 1\% and 10\% phishing ratios, models maintain robust performance, with Grok-3-Beta showing strong zero-shot capabilities. Few-shot learning consistently improves F1 scores, with Grok-3-Beta reaching 0.906 at $\mathcal{E}$=1 under 10\% imbalance. These results demonstrate that prompt-based approaches can effectively detect phishing URLs with minimal examples.

Future work should validate these findings across diverse phishing datasets, explore prompt optimization strategies, and investigate the trade-offs between prompt-based inference and fine-tuned models in operational settings.

\bibliographystyle{plainnat}
\bibliography{reference}

@article{omari2025advanced,
  title={Advanced Phishing Website Detection with SMOTETomek-XGB: Addressing Class Imbalance for Optimal Results},
  author={Omari, Kamal and Oukhatar, Ayoub},
  journal={Procedia Computer Science},
  volume={252},
  pages={289--295},
  year={2025},
  publisher={Elsevier}
}

@article{zara2024phishing,
  title={Phishing website detection using deep learning models},
  author={Zara, Ume and Ayub, Kashif and Khan, Hikmat Ullah and Daud, Ali and Alsahfi, Tariq and Gulzar, Saima},
  journal={IEEE Access},
  year={2024},
  publisher={IEEE}
}

@article{kocyigit2024enhanced,
  title={Enhanced feature selection using genetic algorithm for machine-learning-based phishing URL detection},
  author={Kocyigit, Emre and Korkmaz, Mehmet and Sahingoz, Ozgur Koray and Diri, Banu},
  journal={Applied sciences},
  volume={14},
  number={14},
  pages={6081},
  year={2024},
  publisher={MDPI}
}

@article{ghalechyan2024phishing,
  title={Phishing URL detection with neural networks: an empirical study},
  author={Ghalechyan, Hayk and Israyelyan, Elina and Arakelyan, Avag and Hovhannisyan, Gerasim and Davtyan, Arman},
  journal={Scientific Reports},
  volume={14},
  number={1},
  pages={25134},
  year={2024},
  publisher={Nature Publishing Group UK London}
}

@article{rafsanjani2024enhancing,
  title={Enhancing malicious URL detection: A novel framework leveraging priority coefficient and feature evaluation},
  author={Rafsanjani, Ahmad Sahban and Kamaruddin, Norshaliza Binti and Behjati, Mehran and Aslam, Saad and Sarfaraz, Aaliya and Amphawan, Angela},
  journal={IEEE Access},
  year={2024},
  publisher={IEEE}
}

@article{remya2024effective,
  title={An Effective Detection Approach for Phishing URL Using ResMLP},
  author={Remya, S and Pillai, Manu J and Nair, Kajal K and Subbareddy, Somula Rama and Cho, Yong Yun},
  journal={IEEE Access},
  year={2024},
  publisher={IEEE}
}

@article{mahdaouy2024domurls_bert,
  title={DomURLs\_BERT: Pre-trained BERT-based Model for Malicious Domains and URLs Detection and Classification},
  author={Mahdaouy, Abdelkader El and Lamsiyah, Salima and Idrissi, Meryem Janati and Alami, Hamza and Yartaoui, Zakaria and Berrada, Ismail},
  journal={arXiv preprint arXiv:2409.09143},
  year={2024}
}

@article{liu2025pmanet,
  title={PMANet: Malicious URL detection via post-trained language model guided multi-level feature attention network},
  author={Liu, Ruitong and Wang, Yanbin and Xu, Haitao and Qin, Zhan and Zhang, Fan and Liu, Yiwei and Cao, Zheng},
  journal={Information Fusion},
  volume={113},
  pages={102638},
  year={2025},
  publisher={Elsevier}
}

@inproceedings{ali2025phishurldetect,
  title={PhishURLDetect: A parameter efficient fine-tuning of LLMs using LoRA for detection of phishing URLs},
  author={Ali, Irshad and Subba, Basant},
  booktitle={Proceedings of the 26th International Conference on Distributed Computing and Networking},
  pages={278--279},
  year={2025}
}

@article{tian2025past,
  title={From Past to Present: A Survey of Malicious URL Detection Techniques, Datasets and Code Repositories},
  author={Tian, Ye and Yu, Yanqiu and Sun, Jianguo and Wang, Yanbin},
  journal={arXiv preprint arXiv:2504.16449},
  year={2025}
}

@article{nasution2025benchmarking,
  title={Benchmarking 21 Open-Source Large Language Models for Phishing Link Detection with Prompt Engineering},
  author={Nasution, Arbi Haza and Monika, Winda and Onan, Aytug and Murakami, Yohei},
  journal={Information},
  year={2025},
  publisher={Multidisciplinary Digital Publishing Institute}
}

@article{senanayake2025madonna,
  title={MADONNA: Browser-based malicious domain detection using Optimized Neural Network by leveraging AI and feature analysis},
  author={Senanayake, Janaka and Rajapaksha, Sampath and Yanai, Naoto and Kalutarage, Harsha and Komiya, Chika},
  journal={Computers \& Security},
  volume={152},
  pages={104371},
  year={2025},
  publisher={Elsevier}
}

@article{zhou2025integrated,
  title={An integrated CSPPC and BiLSTM framework for malicious URL detection},
  author={Zhou, Jinyang and Zhang, Kun and Bilal, Anas and Zhou, Yu and Fan, Yukang and Pan, Wenting and Xie, Xin and Peng, Qi},
  journal={Scientific Reports},
  volume={15},
  number={1},
  pages={6659},
  year={2025},
  publisher={Nature Publishing Group UK London}
}

@article{barik2025web,
  title={Web-based phishing URL detection model using deep learning optimization techniques},
  author={Barik, Kousik and Misra, Sanjay and Mohan, Raghini},
  journal={International Journal of Data Science and Analytics},
  pages={1--23},
  year={2025},
  publisher={Springer}
}

@article{lee2024multimodal,
  title={Multimodal large language models for phishing webpage detection and identification},
  author={Lee, Jehyun and Lim, Peiyuan and Hooi, Bryan and Divakaran, Dinil Mon},
  journal={arXiv preprint arXiv:2408.05941},
  year={2024}
}

@inproceedings{devlin2019bert,
  title={Bert: Pre-training of deep bidirectional transformers for language understanding},
  author={Devlin, Jacob and Chang, Ming-Wei and Lee, Kenton and Toutanova, Kristina},
  booktitle={Proceedings of the 2019 conference of the North American chapter of the association for computational linguistics: human language technologies, volume 1 (long and short papers)},
  pages={4171--4186},
  year={2019}
}

@article{radford2018improving,
  title={Improving language understanding by generative pre-training},
  author={Radford, Alec and Narasimhan, Karthik and Salimans, Tim and Sutskever, Ilya and others},
  year={2018},
  publisher={San Francisco, CA, USA}
}

@article{hu2022lora,
  title={Lora: Low-rank adaptation of large language models.},
  author={Hu, Edward J and Shen, Yelong and Wallis, Phillip and Allen-Zhu, Zeyuan and Li, Yuanzhi and Wang, Shean and Wang, Lu and Chen, Weizhu and others},
  journal={ICLR},
  volume={1},
  number={2},
  pages={3},
  year={2022}
}

@article{pedregosa2011scikit,
  title={Scikit-learn: Machine learning in Python},
  author={Pedregosa, Fabian and Varoquaux, Ga{\"e}l and Gramfort, Alexandre and Michel, Vincent and Thirion, Bertrand and Grisel, Olivier and Blondel, Mathieu and Prettenhofer, Peter and Weiss, Ron and Dubourg, Vincent and others},
  journal={the Journal of machine Learning research},
  volume={12},
  pages={2825--2830},
  year={2011},
  publisher={JMLR. org}
}

@article{brown2020language,
  title={Language models are few-shot learners},
  author={Brown, Tom and Mann, Benjamin and Ryder, Nick and Subbiah, Melanie and Kaplan, Jared D and Dhariwal, Prafulla and Neelakantan, Arvind and Shyam, Pranav and Sastry, Girish and Askell, Amanda and others},
  journal={Advances in neural information processing systems},
  volume={33},
  pages={1877--1901},
  year={2020}
}

@misc{ibm2024xforce,

  author       = {{IBM Security}},

  title        = {X-Force Threat Intelligence Index 2024},

  year         = {2024},

  howpublished = {\url{https://www.ibm.com/downloads/documents/us-en/107a02e94948f4ec}},

  note         = {Accessed May 10, 2025}

}

@misc{hoxhunt2024trends,

  author       = {{Hoxhunt}},

  title        = {Phishing Trends Report 2024},

  year         = {2024},

  howpublished = {\url{https://hoxhunt.com/guide/phishing-trends-report}},

  note         = {Accessed May 10, 2025}

}

\end{document}